\documentstyle[12pt,a4wide,bbold,axodraw]{article}

\def\bea{\begin{eqnarray}}
\def\eea{\end{eqnarray}}
\def\be{\begin{equation}}
\def\ee{\end{equation}}
\def\nn{\nonumber}

\def\uni{{\mathbb{1}}}

\newcommand{\eqn}[1]{(\ref{#1})}
\newcommand{\sect}[1]{\setcounter{equation}{0}\section{#1}}

\def\a{\alpha}
\def\b{\beta}

\def\d{\delta}
\def\ve{\varepsilon}
\def\e{\epsilon}

\def\g{\gamma}
\def\G{\Gamma}

\def\m\mu 
\def\n{\nu}

\def\om{\omega}

\def\s{\sigma}

\def\th{\theta}

\def\p{\partial}
\def\tS{\tilde S}

\def\rr{{\rm R}\otimes {\rm R}}
\def\ns{{\rm NS}\otimes {\rm NS}}

\newcommand{\NPB}[3]{{Nucl.\ Phys.} {\bf B#1} (#2) #3}
\newcommand{\JMP}[3]{{Journ.\ Math.\ Phys.} {\bf #1} (#2) #3}
\newcommand{\IJMPA}[3]{{Int.\ J.\ Mod.\ Phys.} {\bf A#1} (#2) #3}
\newcommand{\CMP}[3]{{Commun.\ Math.\ Phys.} {\bf #1} (#2) #3}
\newcommand{\PRD}[3]{{Phys.\ Rev.} {\bf D#1} (#2) #3}
\newcommand{\PLB}[3]{{Phys.\ Lett.} {\bf B#1} (#2) #3}
\newcommand{\JHEP}[3]{{JHEP} {\bf #1} (#2) #3}
\newcommand{\PRL}[3]{{Phys.\ Rev.\ Lett.} {\bf #1} (#2) #3}
\newcommand{\CQG}[3]{{Class.\ Quant.\ Grav.} {\bf #1} (#2) #3}
\newcommand{\T}[1]{``{\em #1}''} 
\newcommand{\hepth}[1]{{\tt hep-th/#1}}
\newcommand{\ft}[2]{{\textstyle\frac{#1}{#2}}\,}
\newcommand{\XX}[2]{\{X^#1,X^#2\}}

\begin{document}

\thispagestyle{empty}
\begin{flushright}
{\small hep-th/0003280}\\
{\small AEI-2000-019}\\[3mm]
\end{flushright}

\vspace{1cm}
\setcounter{footnote}{0}
\begin{center}
{\Large{\bf Vertex Operators for the Supermembrane}
    }\\[14mm]

 {\sc Arundhati Dasgupta, Hermann Nicolai and Jan Plefka}\\[10mm]

{\em Albert-Einstein-Institut}\\
{\em Max-Planck-Institut f\"ur
Gravitationsphysik}\\
{\em Am M\"uhlenberg 1, D-14476 Golm, Germany}\\
{\footnotesize \tt dasgupta,nicolai,plefka@aei-potsdam.mpg.de}\\[7mm]

{\sc Abstract}\\
\end{center}
We derive the vertex operators that are expected to govern the emission 
of the massless $d=11$ supermultiplet from the supermembrane in the
light cone gauge. We demonstrate that they form a 
representation of the supersymmetry algebra and reduce to
the type IIA superstring vertex operators under double dimensional
reduction, as well as to the vertices of the $d=11$ superparticle in
the point-particle limit. As a byproduct, our results can be 
used to derive the corresponding vertex operators for matrix
theory and to describe its linear coupling to an arbitrary $d=11$
supergravity background. Possible applications are discussed.

\vfill
\leftline{{\sc March 2000}}

\newpage
\setcounter{page}{1}

\sect{Introduction}

The fundamental supermembrane \cite{BST} has many features that 
make it an attractive candidate for a fundamental description of 
M Theory at the microscopic level (see e.g. \cite{reviews} for many 
further references). As special limits, it contains the type II
superstrings \cite{superstring} as well as the $d=11$ superparticle 
\cite{superparticle} and is thereby also related to maximal $d=11$ 
supergravity \cite{CJS}. Furthermore, matrix theory can be obtained 
as a regularization of the fundamental supermembrane \cite{dWHN}. 
The theory thus sits atop the main 
contenders for a unified theory of quantum gravity, but actually 
possesses even more degrees of freedom. This is obvious for the 
superparticle, where one retains only the degrees of freedom
corresponding to the $d=11$ supermultiplet, discarding all internal
excitations of the membrane. In the superstring truncation, which 
can be obtained at the kinematical level by a simple procedure called 
double dimensional reduction \cite{DHIS}, one keeps the infinite tower 
of perturbative excited superstring states, but loses the true
M Theory degrees of freedom. However, one still recovers in this way 
both the IIA and IIB superstrings if one keeps the winding
modes and associated BPS multiplets \cite{duality,AdWLN}. Finally, 
maximally supersymmetric matrix theory, which was proposed as a candidate 
for M Theory in the light cone gauge \cite{BFSS,matrix}, does capture the 
non-perturbative degrees of freedom, but only finitely many (and 
misses the winding states of the membrane). At least in the opinion
of the present authors, the successes of the matrix theory proposal
are really rooted in the supermembrane origin of the theory.
In particular, supermembrane theory naturally accounts for all 
aspects related to longitudinal degrees of freedom, which have 
to be guessed in matrix theory because supersymmetric 
Yang Mills theory does not ``know'' about an 11th dimension.

Why is it, then, that supermembrane theory has not gained wider
acceptance, despite all its appealing features? One obvious reason 
is the intrinsic nonlinearity of the theory that 
makes it much harder to deal with than the superstring, and that
has until now blunted all attempts to make meaningful calculations 
{\it at the quantum level} (of course, there is much work on classical 
and semiclassical aspects of the $d=11$ supermembrane, see e.g. 
\cite{solitons}). The supersymmetric $SU(N)$ matrix theory, 
on the other hand, does have the advantage of being rigorously defined 
as a model of quantum mechanics (for finite $N$), and at the same 
time being an intrinsically non-perturbative approximation, but it, 
too, suffers from a host of unsolved problems, especially concerning 
the existence and precise nature of the $N\rightarrow\infty$ limit. 
All matrix theory calculations performed so far are consequently 
limited in scope; for instance, scattering amplitudes have only been
calculated in the eikonal regime where no longitudinal momentum 
transfer is allowed \cite{scattering,SPW}. A recent test of the $R^4$ 
corrections has failed to reproduce the structures predicted by 
string theory \cite{HPSW}.

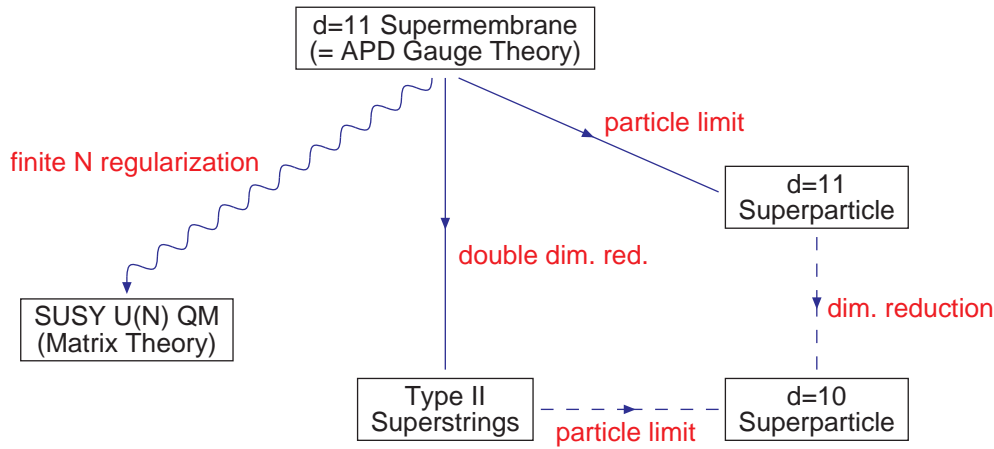
\begin{figure}
\begin{center}
\begin{picture}(280,180)(0,20)
\SetPFont{Helvetica}{10}
\SetScale{1.0}
\B2Text(120,180){d=11 Supermembrane}{(= APD Gauge Theory)}
\B2Text(120,40){Type II}{Superstrings}
\B2Text(0,70){SUSY U(N) QM}{(Matrix Theory)}
\B2Text(260,120){d=11}{Superparticle}
\B2Text(260,40){d=10}{Superparticle}
\SetColor{Blue}
\ArrowLine(120,165)(120,55)
\Photon(115,165)(0,90){3}{10}
\LongArrow(0,90)(0,85)
\ArrowLine(125,165)(223,122)
\DashArrowLine(260,105)(260,55){5}
\DashArrowLine(156,40)(223,40){5}
\SetColor{Red}
\PText(125,100)(0)[lc]{double dim. red.}
\PText(50,135)(0)[rc]{finite N regularization}
\PText(180,150)(0)[lc]{particle limit}
\PText(264,80)(0)[lc]{dim. reduction}
\PText(188,38)(0)[ct]{particle limit}
\end{picture}
\end{center}
\caption{The hierarchy of theories.}
\end{figure}

A further difficulty with supermembrane theory is that we have at 
present very little idea of what the sensible objects are to 
consider and the relevant quantities to compute. This question is 
related to our lack of understanding as to what the fundamental 
supermembrane degrees of freedom really are. Owing to the continuity 
of the supermembrane spectrum \cite{dWLN} (independently realized 
in \cite{Smilga}) there appears to be no analog of the perturbative 
excited superstring states, even though the supermembrane
has far more degrees of freedom. A crucial insight, occasioned by
the matrix proposal, was that the excitations of the theory are
to be associated with multi-particle rather than one-particle
states \cite{BFSS}. The degeneracy of the membrane with regard to 
stringlike deformations suggests a similar picture \cite{HN}.
The only sensible one-particle-like excitations of the theory appear
to be the ones associated with the massless $d=11$ supermultiplet.
However, it does not seem to be possible to set up the usual perturbative 
scheme based on Fock space quantization, or even to assign a definite 
``membrane number'' to a given supermembrane configuration.

In this paper we take a step in the direction of making supermembrane 
theory ``more computable''. By generalizing previous work on 
superstring theory \cite{superstring} and the more recent 
construction of the $d=11$ superparticle vertex operators \cite{GKG}, 
we have succeeded in identifying the supermembrane vertex operators
that are expected to govern the emission of the massless $d=11$
multiplet from the supermembrane. By construction, our vertex operators
contain all previous ones, but they also furnish new information. 
Namely, as a byproduct of the present construction, we are able to 
solve two outstanding and closely related problems of matrix theory: 
the construction of matrix vertex operators and the coupling of the
matrix model to a nontrivial $d=11$ supergravity background in the light 
cone gauge. In this way, we are now in a position to investigate how the 
various presently available results on superstring and matrix model 
amplitudes as well as the non-perturbative results of \cite{GGV,RT,GKG}) 
embed into supermembrane theory. 

The figure displays how the various
theories and their vertex operators are contained in the supermembrane.
The embedding of vertex operators corresponding to the dashed lines
was already studied in \cite{GKG}.

\sect{The supermembrane as a supersymmetric gauge theory of
area preserving diffeomorphisms}

Supermembrane theory was originally formulated as a covariant theory 
coupled to an arbitrary background satisfying the equations
of motion of $d=11$ supergravity \cite{BST}. There are eleven
bosonic target space coordinates $X^M = (X^a, X^\pm)$ (where
indices $a,b,\dots =1,\dots,9$ label the transverse dimensions),
and 32 fermionic fields $\Theta$, which transform as $SO(1,10)$
spinors, but are world volume scalars. All of these fields depend 
on the membrane world volume coordinates $(\tau,\s^1,\s^2)$. 
Like with superstring theory, the supermembrane action simplifies
dramatically when one imposes the light cone gauge  $X^+ = p^+ \tau $ 
and $\Gamma^+ \Theta =0$ (in the following we shall set $p^+=1$ for
simplicity, moreover $\Gamma^\pm=( \Gamma^{10}\pm\Gamma^0)/\sqrt{2}$). 
These conditions reduce the number of 
bosonic degrees of freedom to the nine transverse ones, and halve 
the number of fermionic degrees of freedom to the 16 components 
of an $SO(9)$ spinor $\th$.

An important property of the light cone gauge fixed theory is its 
invariance under a residual infinite dimensional group, the group of
area preserving diffeomorphisms (APDs) \cite{GH} (whose analog for string 
theory simply consists of the constant shifts along the spacelike 
worldsheet coordinate $\s^1$). The canonical constraint associated 
with the APDs is actually necessary to eliminate one further 
bosonic degree of freedom in order to balance the number of bosonic 
and fermionic degrees of freedom on shell, as is required for a 
supersymmetric theory. For any two functions $A(\s^1,\s^2)$ and
$B(\s^1,\s^2)$ on the membrane the APD Lie bracket is given by
\be
\{A,B\} := \ve^{rs} \p_r A \,\p_s B
\ee
where $\p_r :=  \p/\p\s^r$. In this interpretation, one views
the coordinates $(\s^1,\s^2)$ not as providing coordinates 
for the membrane, but rather as a parametrization of the APD Lie
algebra elements.

This residual invariance can be exploited to reformulate the 
light cone supermembrane theory as a supersymmetric gauge theory
of area preserving diffeomorphisms \cite{dWHN}, thereby establishing
the link between the supermembrane and maximally extended supersymmetric
Yang Mills theory. To this aim one introduces (by hand) an APD gauge 
field $\om$, such that in the gauge $\om =0$ one reobtains the original
supermembrane action in the light cone gauge. The resulting theory 
coincides with the dimensional reduction of maximally supersymmetric 
Yang Mills theory {\it to one (time) dimension}, i.e. a model of 
supersymmetric quantum mechanics, but with an infinite dimensional
gauge group (for finite dimensional gauge groups, these models were originally
derived in \cite{SUSYQM}). The APD gauge field $\om$ then coincides 
with the time component of the gauge field of dimensionally reduced 
super-Yang-Mills theory.

The supersymmetric lagrangian of the APD gauge theory reads
\be
{\cal L}= \ft 1 2 (DX^a)^2 - i \theta\, D\theta - \ft 14 \{X^a,X^b\}^2
- i \theta\, \g_a \{ X^a ,\theta\}
\label{m2lagrangian}
\ee
where 
\be
D{\cal O}=\p_0 {\cal O} - \{\omega, {\cal O}\} \, ,
\ee
and where the infinitesimal area preserving diffeomorphisms 
\be
\s^r \rightarrow  \s^r+\epsilon^{rs}\partial_s\xi(\s)
\ee 
act on the fields as $\delta X^a=\{\xi,X^a\}$, 
$\delta \theta=\{\xi,\theta\}$
and $\delta \omega=\partial_0\xi+ \{\xi,\omega\}$.

The lagrangian (\ref{m2lagrangian}) is invariant under the supersymmetry 
variations
\bea
\delta X^a &=& -2 \epsilon\gamma^a\theta \nn\\
\delta \theta &=& i\, DX\cdot \gamma\, \epsilon -\ft i 2 \{X^a,X^b\}\, 
\gamma_{ab}\epsilon +\eta\nn\\
\delta \omega&=& -2 \epsilon \theta
\label{susytrafos}
\eea
As expected from the $d=11$ origin of the model, there are
still 32 supersymmetry parameters. These are split into two  
16-component $SO(9)$ spinors $\eta$ and $\e$. Following established
usage, we will refer to them as {\it linear supersymmetry} 
(parametrized by $\eta$) and {\it nonlinear supersymmetry} 
(parametrized by $\e$) transformations, respectively. The linear 
supersymmetry transformations obviously affect only the zero modes. 

The equations of motion that follow from the above action are
\bea
0&=& D^2 X^a - \{\, \{X^a,X^b\}, X^b\} - i \{\theta, \gamma^a\theta\} 
\label{eomx}\\
0&=& D\theta +\{\gamma\cdot X,\theta\} \label{eomth}\\
0&=& \{DX^a,X^a\} - i \{\theta,\theta\} 
\label{eomomega}
\eea
The last of these equations, obtained by varying the gauge field $\om$,
is the constraint associated with the APD gauge invariance
on the membrane.

As shown in \cite{dWHN}, the above model can be approximated by
a supersymmetric $SU(N)$ matrix model, such that the full theory is 
(formally) recovered by taking the limit $N\rightarrow\infty$.
The essential ingredient here is the result that the group of APDs 
can be approximated by $SU(N)$. This statement, first established for
spherical membranes in \cite{GH} and for toroidal ones in \cite{Hoppe,dWMN},
actually holds for membranes of arbitrary topology \cite{B}.
The prescription for obtaining a matrix model from the above lagrangian
is simple: just replace the target space fields by $SU(N)$ matrices 
according to
\bea
X^a(\tau, \s^1,\s^2) &\longrightarrow&  {\bf X}^a_{mn}(\tau) \equiv
\sum_A X^{aA}(\tau) \,  Y^A_{mn} \nn \\
\th_\a(\tau ,\s^1,\s^2) &\longrightarrow&  {\bf \th}^\a_{mn}(\tau) \equiv 
\sum_A \th^A_\a (\tau) \, Y^A_{mn}
\label{mtrans}
\eea
with $m,n=1,\ldots, N$ labeling the entries of the $N\times N$
hermitian matrices ${\bf X}^a$ and $\theta^\a$, and a (hermitian and
orthonormal) basis $\{ Y^A| A=1,\dots,N^2-1 \}$ of the $SU(N)$ Lie algebra. 
Furthermore, the APD Lie bracket gets replaced by a matrix commutator
$\{.\, , .\} \rightarrow i\, [.\, , .]$. This is all that 
is needed to get the matrix model, proposed as a candidate 
for a microscopic description of M Theory in the light cone gauge \cite{BFSS}. 
We will return to matrix theory in section 5, to show how our results
can be exploited to derive vertex operators for the matrix model,
and to couple the matrix model to a nontrivial $d=11$ background.

An equally important property of the supermembrane is that
it contains the superstring as a special truncation. The embedding
is achieved by identifying the membrane target space coordinate
$X^{9}$ with the world volume coordinate $\s^2$, a procedure 
called ``double dimensional reduction'' \cite{DHIS}. Setting
$X^{9} = \s^2$ and letting all other fields only depend on $\s^1$, 
with $i,j$ labeling the first eight transverse directions,
the action \eqn{m2lagrangian} collapses to\footnote{Since
$\s^2 \in [0,R)$, there are also the winding modes associated
with the compactification on the circle.}
\be\label{GSaction}
{\cal L}_{DDR} = \ft12 (\p_0 X^i)^2 - \ft12 (\p_1 X^i)^2 - 
           i\th\p_0\th+i\th\g^9\p_1\th
\ee
which is just the Green Schwarz light cone lagrangian of
the IIA superstring\footnote{For the $SO(9)$ Clifford algebra,
we choose the following representation
$$
\g^9 = 
\left(\begin{array}{cc} \uni& 0 \\
                             0 & -\uni
\end{array}\right)
\qquad
\g^i =
\left(\begin{array}{cc} 0 & \Gamma^i \\
                      \bar\Gamma^i  & 0
\end{array}\right)
$$
where $\G^i_{\a\dot\b}$ and $\bar\G^i_{\dot\a \b}$ are the standard
$SO(8)$ $\G$-matrices.}. As we will see, the superstring vertex 
operators can be recovered from those of the supermembrane by an
analogous procedure.

\sect{The Vertex Operators}
The massless states of the supermembrane (and the supersymmetric
matrix model) are expected to yield a massless multiplet
of $d=11$ supergravity, containing the graviton, the three-form
gauge potential and the gravitino (see \cite{massless} for
progress in establishing the existence of such states).
We are therefore interested in constructing candidates 
for vertex operators that would describe the emission of 
these massless states from the supermembrane (due to the continuity
of the supermembrane mass spectrum, there appear to be no
discrete excited supermembrane states). Clearly, an essential 
consistency requirement for such operators is that they should 
coincide with the corresponding ones of the $d=11$ superparticle
\cite{GKG}, as well as with the full superstring vertex operators
upon double dimensional reduction. We note in passing that the leading
$\theta$ contribution to the (covariant) gravitino vertex operator 
has already been used in computations of membrane instanton effects 
in \cite{m2inst} .

To arrive  at closed expressions for the vertex operators, 
we follow the strategy that was already successfully employed 
in the construction of superstring vertex operators \cite{superstring}, 
and more recently the construction of vertex operators for 
the $d=11$ superparticle \cite{GKG}. Namely, one exploits the 
fact that under the above supersymmetries the vertex operators 
should vary into one another, such that the transformations 
can be thrown onto the corresponding variations of the 
polarizations as they follow from $d=11$ supergravity.
Schematically, we thus have
\bea
\d V_h &=& V_{\d\psi[h]} \nn \\
\d V_C &=& V_{\d\psi[C]} \nn \\
\d V_\psi &=& V_{\d h} + V_{\d C}
\eea
up to total derivatives (while the total derivatives in \cite{GKG}
were always derivatives w.r.t. to time, they here appear both as
$D(\dots)$ and $\{.,.\}$). The variations to be performed on the l.h.s.
of these expressions are the ones of the supersymmetric 
APD gauge theory given
in \eqn{susytrafos} above, whereas the variations on the r.h.s. are those
induced by $d=11$ supergravity on the various polarizations. By 
$\d \psi [h]$ and $\d \psi [C]$ we here designate the terms in the 
gravitino variation depending on the graviton and three-form 
polarizations $h$ and $C$, respectively. A detailed explanation 
of the general procedure can for instance be found in \cite{GKG}.

An alternative route to arrive at our results would be to start from 
the covariant $d=11$ supermembrane \cite{BST}, whose background 
coupling is explicitly known in {\it superspace}. This approach
would yield the fully covariant vertices, which should then reduce
to the vertices presented above in the light cone gauge. However,
obtaining the component form of this action by constructing the superspace 
vielbein and tensor gauge field in terms of the component fields 
to all orders in $\theta$ appears to be a prohibitively difficult task: 
to date the expansion is only known up to order $\theta^2$ for general 
backgrounds \cite{dWPP}(incidentally, the fully covariant vertex operators 
are not even known for the GS superstring or the superparticle). 
The light-cone approach adopted here proves to be far more efficient
because the expansion in $\theta$ already terminates at order five -- as 
opposed to order 32 for the covariant expressions. This demonstrates 
again the drastic simplification of the fermion sector in the light-cone 
gauge, already seen for the flat background action. 

Let us now present the results, and then comment on their
derivation and the various consistency checks which we have
performed to ascertain the correctness of these expressions. All vertex 
operators come with a factor $\exp(-ik\cdot X)\,\exp(i k^-\tau)$, 
where $k_a$ is the (transverse) momentum of the state emitted.
Following standard practice in string theory \cite{GS82}, we will set
$k^+ \equiv k_- = 0$ in order to avoid the appearance of the 
longitudinal target space coordinate $X^-(\tau,\s^1,\s^2)$ in the 
exponential\footnote{We are aware that this choice of frame is somewhat 
questionable, although widely adopted: with it, the transverse 
momentum components must become complex. In addition, there are 
inverse factors of $1/k^+$ in some of the compensating transformations;
fortunately, these drop out due to the gauge invariance of 
the vertices.}. Furthermore we shall often disregard the extra 
factor $\exp (i k^-\tau)$ in our considerations, except in 
those places where it gives extra contributions from integrating
by parts the time derivative operator $D$.

The vertex operators are contracted with the polarizations 
corresponding to the massless states of $d=11$ supergravity. 
Choosing the gauge conditions $h_{a-} = h_{--} = h_{+-} = 0 \, , \,
C_{ab-}= C_{a+-} = 0  \, , \, \psi_- = \tilde\psi_- =0$, and 
splitting the remaining polarizations according to their 
longitudinal content, we have 
\bea
{\rm graviton:}  \qquad && (h_{ab} , h_{a+}, h_{++}) \nn \\
{\rm three-form:}    \qquad && (C_{abc} , C_{ab+}) \nn \\
{\rm gravitino:} \qquad && (\psi_a , \psi_+; \tilde\psi_a, \tilde\psi_+)
\eea
Note that again we have 32 spinor components in accordance with the
$d=11$ origin of the model, namely 16 components for $\psi$ and 
$\tilde\psi$ each. We will also use the gauge invariant combinations
\be
F_{abcd} := 4 k_{[a} C_{bcd]} \qquad
F_{abc+} := 3 k_{[a} C_{bc]+} + k_+ C_{abc}
\ee
The polarizations are subject to the 
following physical state constraints \cite{GKG}: 
\bea
&k^a h_{ab} = h_{aa} = 0 = k^a C_{abc}&\nn \\
&\g^a \,\tilde\psi_a = \g^b k_b\, \tilde\psi_a = k^a\,\tilde\psi_a
= 0 =k^a\,\psi_a &\nn\\
&\g^a\, \psi_a=\tilde\psi_+ \qquad k^b \g_b\, \psi_a= k^-\, \tilde\psi_a
\label{sugraconst}
\eea
For the bilinears in the sixteen component real spinors $\th$ we 
introduce the notation
\be
R^{abc}=\ft 1 {12}\, \th\g^{abc}\th \qquad
R^{ab}=\ft 1 {4}\, \th\g^{ab}\th \, .
\ee
The key $SO(9)$ Fierz identity for $\th$ reads
\be
\th_\alpha\, \th_\beta= \ft 1 2 \delta_{\alpha\beta}\, \d^{(2)}(0)
+\ft 1 {32}\, \g_{\alpha\beta}^{ab}\, \th\g_{ab}\th
+\ft 1 {96}\, \g_{\alpha\beta}^{abc}\, \th\g_{abc}\th \, ,
\ee
The singular $\delta_{\alpha\beta}\, \d^{(2)}(0)$ term here arises if one 
assumes the standard canonical anticommutation relations for the 
fermionic operators $\th_\alpha$. Fortunately, however, this term 
drops out in all the manipulations performed in this work and is thus
irrelevant to our final expressions.

Let us now state the main results of this paper and describe its
derivation in the next chapter. The graviton vertex operator is given by
\bea
V_h&=& h_{ab}\, \Bigl [ DX^a\, DX^b - \{X^a,X^c\}\,\{X^b,X^c\} - i
\theta\gamma^a\,\{X^b,\theta\} \nn\\
&& -2 DX^a\, R^{bc}\, k_c - 6 \{X^a,X^c\}\, R^{bcd}\, k_d 
+ 2 R^{ac}\,R^{bd}\, k_c\, k_d \Bigr ]\, e^{-i k\cdot X} \label{Vh}\\
V_{h_+} &=& -2 h_{a+} \, (DX^a - R^{ab} k_b) e^{-i k\cdot X} \label{Vh+}\\
V_{h_{++}} &=&  h_{++}\, e^{-i k\cdot X}
\label{Vh++}
\eea
For the vertex operator corresponding to the three-form potential, we find
\bea
V_C &=& -C_{abc}\, DX^a\, \XX{b}{c}\, e^{-i k\cdot X} + F_{abcd}\Bigl[
(DX^a-\ft 23 R^{ae}\, k_e)\, R^{bcd} \nn\\
&& -\ft 12 \XX{a}{b}\, R^{cd} -\ft 1 {96}\, \XX ef\, \theta \gamma^{abcdef}
\theta \Bigr ]\, e^{-i k\cdot X} \label{VC}\\
V_{C_+} &=& C_{ab+} (\XX a b +3 R^{abc}\, k_c) e^{-i k\cdot X}
\label{VC+}
\eea
Finally, for the gravitino vertex operators, we obtain
\bea
V_\Psi &=& 
\psi_a\, \Bigl [ \, \Bigl (DX^a-2 R^{ab}\, k_b +\gamma_c\,
\XX ca\, \Bigr ) \theta \Bigr ]\, e^{-i k\cdot X}\nn\\
&&+ \tilde\psi_a\, \Bigl [ \gamma\cdot DX\,  \Bigl (DX^a- 2R^{ab}\, k_b
+\gamma_{c} \XX ca \Bigr )\,\theta \nn\\
&&+\ft 12 \gamma_{bc}\, \XX b c \, ( DX^a- \XX a d \, \gamma^d\, )\theta
+ 8  \gamma_b\theta\, \XX b c \, R^{cad}\, k_d \nn\\
&&+ \ft 5 3 \gamma_{bc}\theta\, \XX b c R^{ad}\,k_d
+ \ft 4 3 \gamma_{bc}\theta\, \Bigl ( \XX a b\, R^{cd} + \XX c d R^{ab}
\Bigr ) k_d \nn\\
&&
+\ft 2 3 i \,\Bigl ( \gamma_b\theta\, \{X^a,\theta\}\gamma^b\theta
-\theta\, \{X^a,\theta\}\theta \,\Bigr ) 
+ \ft 8 9 \gamma^b\theta\, R^{ac}\, R^{bd}\, k_c\, k_d
\,\Bigr ] \, e^{-i k\cdot X} \label{VPsi}\\
V_{\Psi_+} &=& -\left[\psi_+  \th + \tilde\psi_+ \left( \g^a DX^a +
             \ft12 \g^{ab} \{ X^a,X^b\} \right) \th \right] e^{-i k\cdot X}
\label{VPsi-}
\eea
Concerning the above vertices one should keep in mind that relaxing
the frame choice $k^+ =0$, we would have to cope with extra terms
involving the longitudinal component $k^+ X^-$ not only in the 
exponential, but also in the prefactors multiplying the exponential.
Secondly, when passing to the quantum theory we must be prepared 
to modify the vertices by extra ``renormalizations'' as would
be the case for composite operators in any interacting quantum 
field theory (such as QCD). However, such modifications are very 
tightly constrained in that they must not only preserve the symmetry 
properties to be discussed below, but also reduce to the standard 
normal-ordering prescription in the superstring limit.

\sect{Consistency Checks}
The complete expressions given above were arrived at by exploiting a 
number of constraints and consistency requirements. There are 
altogether four of these, which follow from $(i)$ gauge invariance, 
$(ii)$ dimensional reduction, $(iii)$ linear supersymmetry, and  
$(iv)$ nonlinear supersymmetry. We will now discuss these in turn.
A further (and quite tedious) check, which we have not performed,
would be to verify the covariance of the vertices under Lorentz 
boosts in eleven dimensions, using the supermembrane boost generators 
constructed in \cite{dWMN}.

\subsection{Gauge invariance}
Gauge invariance of the vertices requires that they be left
unchanged under the following transformations, 
\bea \d h_{ab} &=& k_{(a} \xi_{b)}  \qquad
     \d h_{a+}  =  \ft 12 (k_{a} \xi_{+} + k_+ \xi_a ) \qquad
     \d h_{++}  =   k_{+} \xi_{+}    \\
     \d C_{abc}&=& 3 k_{[a} \xi_{bc]} \qquad 
     \d C_{ab+} = 2 k_{[a} \xi_{b]+} + k_+ \xi_{ab} \\
     \d \psi_a &=& k_a \e \qquad \d \tilde\psi_a = k_a \eta \nn \\
     \d \psi_+ &=& k_+ \e \qquad \d \tilde\psi_+ = k_+ \eta
\eea 
which are induced on the polarization tensors by the corresponding
gauge symmetries of $d=11$ supergravity. The transformations listed, 
respectively, correspond to (linearized) coordinate transformations
(with parameter $\xi^a$), to tensor gauge transformations (with
parameter $\xi_{ab}=-\xi_{ba}$), and to the inhomogeneous (field
independent) part of the supersymmetry transformations. Gauge invariance
holds only on-shell, because in order to establish it, we will have to 
make use of the equations of motion \eqn{eomx}, \eqn{eomth} and 
\eqn{eomomega}.

The invariance under tensor gauge transformations is manifest in the 
transverse sector, except for the first term in $V_C$, which transforms 
as
\bea
\d_\xi V_C &=& \Bigl [-\xi_{bc}\, k\cdot DX\, \XX{b}{c}- 2\xi_{bc}
\, DX^b\, \{X^c,k\cdot X\}\Bigr ]\, e^{-i k\cdot X + i k^-\tau}\nn\\
&=& - k^-\, \xi_{bc}\,\XX{b}{c}\, e^{-i k\cdot X + i k^-\tau}
\eea
upon partial integration. This precisely cancels the variation
of $V_{C_+}$ as
\be
\d_\xi V_{C_+} = 2i\, \xi_{b+}\, \{ e^{-i k\cdot X + i k^-\tau}, X^b\}
+ k^-\, \xi_{ab}\, \XX{a}{b}\, e^{-i k\cdot X + i k^-\tau}\, ,
\label{da}
\ee
where the first term in \eqn{da} is a total derivative.

The graviton vertex requires a little more work: replacing $h_{ab}$
by $k_{(a} \xi_{b)}$ in \eqn{Vh} we see that several terms drop out by
antisymmetry. For the remaining ones, we get
\bea
\d_\xi V_h &=& \Big[ D(k\cdot X) ( D(\xi \cdot X) - R^{ab} \xi_a k_b )
       - \{ k\cdot X ,X^c\} \{\xi\cdot X, X^c\}  \nn\\
     && 3\{ k\cdot X, X^c\} R^{abc}\xi_a k_b - 
        \ft{i}2 \th \xi_a\g^a \{ k\cdot X , \th \} -
        \ft{i}2 \th k_a\g^a \{\xi\cdot X , \th \}
        \Big] e^{-i k\cdot X}
\eea
Next we integrate by parts the terms involving $k\cdot X$; this yields
\bea
\d_\xi V_h &=& i e^{-i k\cdot X} \Big[ - D^2 (\xi\cdot X) +
   \{X^c,\{X^c, \xi\cdot X \}\} +\ft i 2\{ \th , \xi_a \g^a\th \} \nn\\
&&+\ft 1 2 D\th\g^{ab}\th\, \xi_a\, k_b -\ft 1 2 \th k\cdot \g \, \{
\xi\cdot X, \th\}+\ft 1 2 \, \{X^c,\th\}\g^{abc}\th\, \xi_a\, k_b\Big]\nn\\
&&+ k^-\, ( D(\xi \cdot X) - R^{ab} \xi_a k_b )\, e^{-i k\cdot X}
 \, .
\eea
The terms in the first two lines vanish by making use of the 
equations of motion of $X^a$ \eqn{eomx} 
and $\th$ \eqn{eomth}, whereas the last term is seen to cancel
with the gauge transformations of the longitudinal graviton
vertices
\bea
\delta_\xi V_{h_+}&=& -\Big [\xi^-\, k^- +
k^-\, ( D(\xi \cdot X) - R^{ab} \xi_a k_b )\Big ]
\, e^{-i k\cdot X} \\
\delta_\xi V_{h_{++}}&=& \xi^-\, k^-\, e^{-i k\cdot X} \, .
\eea

For the gravitino vertex again several terms drop out by antisymmetry,
and we are left with
\bea
\d_\xi V_\Psi &=& -i\eta e^{-i k\cdot X} \Big[D\th +\g^a\{ X^a,\th \}\Big]
     +k^-\, \eta\theta\, e^{-i k\cdot X} \nn\\
    && + \e e^{-i k\cdot X} [\g^a DX^a \left( D(k\cdot X) +
                     \g^b \{ X^b , k\cdot X \} \right) \th    \nn \\
    && \qquad + \ft12 \g^{ab} \{ X^a, X^b \} \left( D(k\cdot X) -
          \{k\cdot X , X^c \} \g^c \right) \th \Big]    \nn \\
    && + \ft23 i  e^{-i k\cdot X} \Big[ \e\th \, \th \{k\cdot X , \th \}
      -   \e \g^a \th \, \th \g^a \{k\cdot X , \th \} \Big]
\eea
The first line is just the fermionic equation of motion plus a term
that cancels against the gauge transformation of $V_{\Psi_+}$. The remaining
terms can likewise be shown to vanish on shell after some integrations
by part, and use of the Fierz identity:
\be
\e\g^a\{\th\, ,\,\th\}\th -\e\{\th\, ,\,\th\}\th =
\ft 12 \e\th\, \{\th  , \th\} -\ft 1 2 \e\g_a\th\, \{\th  , \g^a\th\}
\, .
\ee

\subsection{Reductions} 
There are two reductions which provide stringent consistency checks.
The first arises from the comparison of our vertex operators with
those of the superparticle recently determined in \cite{GKG}.
In this truncation one stays in eleven dimensions, but discards all
internal degrees of freedom, such that the variables $(X^a,\th)$
no longer depend on the coordinates $(\s^1,\s^2)$, but only on $\tau$.
Accordingly, one simply drops the terms involving the APD Lie bracket
$\{.,.\}$ in all expressions. Although this looks like a rather trivial 
truncation, it still yields a good deal of the information required; 
in particular, quartic and quintic fermionic terms are not affected 
by it at all, as they are independent of $X^a$. This allows us to 
take over the pertinent expressions from \cite{GKG} and 
thereby to fix many terms without further ado. 

To check the agreement of our vertices with those of superstring theory
(which are also listed in \cite{GKG}) after double dimensional reduction 
is more subtle, not least because some ``obvious'' guesses turn out to 
be incorrect. In this truncation one retains the infinite tower of
(perturbative) massive superstring states together with the BPS states
(the winding states of the membrane), but $d=11$ covariance is 
lost. Demanding the doubly reduced vertices to agree with those of
superstring theory then fixes the terms involving the APD Lie brackets,
which cannot be determined from the superparticle vertex operators. 
It is most remarkable that, despite the absence of any factorization in 
eleven dimensions, our vertices do factorize in precisely the required 
way after dimensional reduction. Furthermore, they combine the 
contributions originating from the $\rr$ and the $\ns$ sectors, which 
superstring theory treats separately, into unified expressions.
 
As already mentioned, upon double dimensional reduction, the 
APD brackets either vanish, or become derivatives w.r.t. to the
remaining string worldsheet coordinate $\s\equiv\s^1$, such that
\be
\{ X^i , X^j \} = 0 \qquad \qquad \{ X^i , X^9 \} = \p_1 X^i
\ee
Adopting the gauge $\om =0$, we must then regroup all terms containing 
derivatives $\p_0$ and $\p_1$ in such a way that the derivatives appear 
only in the left- or right-moving combinations $\p_\pm \equiv \p_0 \pm \p_1$,
as required by consistency. The $SO(9)$ spinors $\th$ must be decomposed 
into $SO(8)$ spinors according to
\be
\th (\tau,\s) = 
\left(\begin{array}{cc} S_\a (\tau,\s)\\
                        \tilde S_{\dot\a} (\tau,\s)
\end{array}\right)
\ee
{} From the equations of motion \eqn{eomth}, or from the reduced action
(\ref{GSaction}), it immediately follows that
$\p_- S = \p_+ \tS = 0$. Therefore, in the reduction the spinor
$\th$ decomposes into the the left- and right-moving free fermions 
of IIA superstring theory. It is easy to see that
\bea
R^{ij}  &=& \ft14 S \G^{ij} S  + \ft14 \tS \G^{ij} \tS  \qquad \qquad
            R^{i9} = \ft12 \tS \G^i S     \\
R^{ijk} &=& \ft16 S \G^{ijk} \tS   \qquad \qquad
    R^{ij9} = \ft1{12} S \G^{ij} S - \ft1{12} \tS \G^{ij} \tS
\eea
in terms of $SO(8)$ spinors. Let us emphasize once more that 
the superparticle reduction ensures that quartic and 
quintic fermionic terms work by themselves, so the tests 
performed below concern only terms containing the APD bracket.

{}For the $d=10$ graviton $h_{ij}$, an $\ns$ field, the double dimensional
reduction gives
\bea
(V_h)_{DDR} &=& h_{ij} \Big[ \p_0 X^i \p_0 X^j -\p_1 X^i \p_1 X^j
           - \ft12 \p_0 X^i (S\G^{jm} S + \tS \G^{jm} \tS ) k_m  \nn \\
   && \qquad + \ft12 \p_1 X^i (S\G^{jm} S - \tS \G^{jm} \tS ) k_m 
   + \ft14 S \G^{im} S \, \tS \G^{jn} \tS k_m k_n \Big] e^{-i k\cdot X} \nn \\
  &=& h_{ij} \Big( \p_+ X^i - \ft12 S\G^{im} S k_m \Big)
        \Big( \p_- X^j - \ft12 \tS\G^{jn} \tS k_n \Big) e^{-i k\cdot X}
\eea
This is the desired result, see e.g. section 4.1 of \cite{GKG}. For the 
$R\otimes R$ vector field $h_{i9}$, we obtain
\be
h_{i9} \Bigl[ -i\th \g^i \p_1 \th + 2 \p_0 X^i R^{m9} k_m 
  + \p_1 X^j R^{ijm} k_m +2 R^{im} R^{9n} k_m k_n \Bigr] e^{-i k\cdot X}
\ee
Again the quartic terms are easily seen to agree. To get rid of 
the derivatives on $\th$, which are absent in the superstring 
vertices, we make use of the superstring equations of motion 
$\p_1 S = \p_0 S$ and $\p_1 \tS = - \p_0 \tS$, and integrate the 
resulting expression by parts. After a little algebra we arrive at
the desired result:
\be
k_{i} h_{j9} \Bigl[ S \G^{ij} \G^{k} \tS \, \p_- X^k -
       S \G^{k} \G^{ij} \tS \, \p_+ X^k \Bigr]  e^{-i k\cdot X}
\ee

The three-form $C_{abc}$ gives rise to the $\rr$ field $C_{ijk}$ and 
the $\ns$ field $C_{ij9}$ in the reduction to ten dimensions, and the 
corresponding vertices must again be checked separately. Dimensional 
reduction of (\ref{VC}) yields 
\bea
(V_C)_{DDR} &=& -C_{ij9}\, \p_0 X^i \p_1 X^j \, e^{-i k\cdot X} \nn \\
&&  +  F_{ijk9}\Bigl[ \ft34 \big(\p_0 X^i - \ft 23 R^{im} k_m\big) R^{jk9} 
                     -\ft16 R^{9m} R^{ijk} k_m
                 - \ft14\p_1 X^i R^{jk} \Bigr]  e^{-i k\cdot X} \nn\\
&& + F_{ijkl}\Bigl[ \big(\p_0 X^i -\ft 23 R^{im}k_m \big) R^{jkl} -
     \ft1{48} \p_1 X^m \, \th \g^{ijklm9} \th \Bigr]  e^{-i k\cdot X}
\eea
The superstring vertices involving the $\rr$ field $C_{ijk}$ can
be deduced from the formulas listed in section (4.1) of \cite{GKG}. 
They are given by (dropping the quartic fermion terms)
\bea
&& \ft 1 {48}F_{ijkl} \Bigl[ S \G^{ijkl} \G^m \tS\, \p_- X^m +
              \tS \G^{ijkl} \G^m S \,\p_+ X^m \Bigr] e^{-i k\cdot X}  \nn \\
&& =  F_{ijkl} \Bigl[ \ft1 6 \, \p_0 X^l\,  S \G^{ijk} \tS -
               \ft 1 {24}\, 
\p_1 X^m \, S \G^{ijklm} \tS \Bigr] e^{-i k\cdot X}
\eea
which indeed agrees with the result derived before. The agreement 
for the $\ns$ vertex involving $C_{ij9}$ is verified similarly.

More work is required to check the gravitino vertex. Most of the terms 
can be guessed correctly by making the ``obvious'' substitutions, such as 
\be
\left(\begin{array}{cc}\p_\pm X^i \,(\G^i \tS)_{\a} \\
                        \p_\mp X^i \, (\bar\G^i S)_{\dot\a}
\end{array}\right)
\longrightarrow 
\Big( \g^a DX^a \pm \ft12 \g^{ab} \{ X^a,X^b \} \Big) \th
\ee
The substitutions for the terms cubic in $\theta$ and containing
an APD bracket are more tricky. Under double dimensional reduction
\bea
6 \tilde\psi_a \g_b \th \, R^{acd} k_c \XX b d &\rightarrow &
\ft12 \big(\tilde\psi_{i\a} \G^j_{\a\dot\b}\tS_{\dot\b} +
      \tilde\psi_{i\dot\a}\G^j_{\dot\a \b} S_{\a}\big)
      \big( S\G^{im} S - \tS \G^{im} \tS \big) k_m \, \p_1 X^j   \nn\\
&& + \big(\tilde\psi_{i\a} S_\a - \tilde\psi_{i\dot\a} \tS_{\dot\a}\big)\,
      S \G^{ijm} \tS k_m \, \p_1 X^j    
\eea
Only the terms on the first line of the r.h.s. agree with the 
corresponding ones for the superstring. To eliminate the unwanted
terms, we must add two further terms to the gravitino vertex, viz.
\be
\tilde\psi_a \g_{bc} \th \, R^{ad} k_d \XX b c
\ee
and
\be
\tilde\psi_a \g_{bc} \th \Big( R^{cd} \XX a b  
                    + R^{ab} \XX c d \Big) k_d
\ee
and, by a judicious choice of coefficients try to cancel them. 
This is indeed possible, if one makes use of the following 
$SO(8)$ Fierz identities 
\bea
\tilde\psi_{i\a} S_\a \, S \G^{imj} \tS k_m \, \p_1 X^j 
&=& -\ft12 \tilde\psi_{i\a} \G^j_{\a\dot\b} \tS_{\dot\b}
           \, S\G^{im} S k_m \, \p_1 X^j \nn \\
&& + \ft14  \tilde\psi_{i\a} \G^j_{\a\dot\b} \tS_{\dot\b}
     \Big( S \G^{ij} S \, k\cdot \p_1 X + S\G^{jm} S k_m \, \p_1 X^i \Big)
\eea
and
\be
  \tilde\psi_{i \a} \G^j_{\a\dot\b} \tS_{\dot\b} \,
      \tS \G^{im} \tS k_m \, \p_1 X^j =
 \ft13 \tilde\psi_{i \a} \G^j_{\a\dot\b} \tS_{\dot\b} \,
 \Big( \tS \G^{ij} \tS \, k\cdot \p_1 X + \tS\G^{jm} \tS k_m 
  \, \p_1 X^i \Big)
\ee
To summarize: the comparison with the $d=11$ superparticle and 
$d=10$ superstring vertices constrains the possible terms so tightly 
that we are left with unique expressions for the supermembrane vertex
operators. The final test is then provided by supersymmetry. 

\subsection{Linear Supersymmetry}
The first consistency check under supersymmetry involves the variation
of the vertex operators \eqn{Vh}, \eqn{VC} and \eqn{VPsi} under the
linear transformations 
\be
\delta X^a=\delta \omega =0 \qquad \mbox{and} \qquad \delta \theta = \eta
\label{linsusy}
\ee
which should induce the homogenous supergravity variations 
(neglecting longitudinal polarizations) \cite{GKG}
\bea
&\delta h_{ab}  = -\tilde\psi_{(a}\g_{b)}\eta  \qquad
\delta h_{a+} = -\ft 1 {\sqrt 2}\, \psi_a\eta \label{dhab}&\\
&\delta C_{abc} = \ft 32 \tilde\psi_{[a}\g_{bc]}\eta \qquad
\delta C_{ab+} = \sqrt 2\, \psi_{[a}\g_{b]}\eta \label{dCabc}&\\
&\delta \psi_{a} = k_{b}\, h_{ca}\, \g^{bc}\eta + \ft {1}{72}
(\g_{a}{}^{bcde}\, F_{bcde}- 8 \g^{bcd}\,F_{abcd})\, \eta \qquad
\delta \tilde\psi_+ =-\ft {\sqrt{2}}{72} \g^{abcd}\eta\, F_{abcd}&\\
&\delta \psi_+ = 
\delta \tilde\psi_{a} =0=\delta h_{++}&\nn
\eea
of the polarizations. As before we work in the kinematical sector 
where $k^+=0$.

Performing the variation \eqn{linsusy} on the transverse graviton
vertex \eqn{Vh} yields
\bea
\delta V_h&=& k_b\, h_{ca} \eta\g^{bc}\,\Bigl [ DX^a-2 R^{ad}\,k_d -\g^d\,
\XX a d\, )\th \Bigr ]\, e^{-ik\cdot X} \nn \\
&& - h_{ab} \Bigl [ \{ X^a,k\cdot X\}\, \eta\g^b\th  + i \eta\g^a\{X^b,\th\}\,
  \Bigr ]\, e^{-ik\cdot X}\nn\\
&=& -V_{\delta \psi[h]}
\eea
where the two terms in the second line cancel via a partial integration.

Next we turn to the transverse 3-from vertex whose variation yields
\bea
\delta V_C&=& F_{abcd}\, \Bigl [ \ft 16 (DX^a-\ft 2 3 R^{ae}\, k_e)
\, \eta\g^{bcd}\th - \ft 1 {36} \eta\g^{ae}\th\, \th\g^{bcd}\th \, k_e
\nn\\&&
-\ft 1 4 \XX a b \, \eta\g^{cd}\th -\ft 1 {48}\, \XX e f \, 
\eta\g^{abcdef}\th\, \Bigr] \, e^{-ik\cdot X}\nn\\
&=&
F_{abcd}\, \Bigl [ \ft 16 DX^a\, \eta\g^{bcd}\th
-\ft 1 4 \XX a b \, \eta\g^{cd}\th -\ft 1 {48}\, \XX e f \, 
\eta\g^{abcdef}\th\, \Bigr ] e^{-ik\cdot X}\nn\\
&&-\ft 1 {36}\, R^{af}\, k_f\, \Bigl [ \eta\g_a{}^{bcde}\th\, F_{bcde}
+ 8\eta\g^{bcd}\th\, F_{abcd}\, \Bigr ]\, e^{-ik\cdot X} 
\label{c1}
\eea 
where we have made use of the Fierz identity
\be
F_{abcd}\, \eta \g^{ae}\th\, \th \g^{bcd}\th k_e=
-F_{abcd}\Bigl [ \eta\g^{abc}\th\, \th\g^{de}\th\, k_e +\ft 1 4
\th\g^{abcde}\eta\, \th\g^{ef}\th\, k_f\Bigr ]
\ee
on the terms of order $\th^3$. This result is to be compared with
the vertex operators of the varied gravitino polarizations 
\bea
V_{\delta \psi_a}+ \ft{1}{\sqrt{2}}\, V_{\delta\tilde\psi_+} &=&
\ft 1{72}\Bigl ( \eta\g_a{}^{bcde}\, F_{bcde} + 8 \eta\g^{bcd}\th\, F_{abcd}
\Bigr )\, \Bigl ( DX^a-2R^{ab}\,k_b -\g^c\, \XX a c \Bigr )\th
\nn\\
&&-\ft 1{72} F_{abcd}\, \eta\g^{abcd}\, \Bigr 
(\gamma\cdot DX+\ft 1 2 \g^{ab}\,
\XX a b\Bigl )\th\, e^{-ik\cdot X} 
\eea
which is easily shown to equal \eqn{c1}.

Finally we examine the linear supersymmetry variation of the gravitino
vertex, which due to its size and the required heavy use of 
Fierz rearrangements in the computation is considerably more involved.

The variation of the $\psi_a$ vertex yields
\be
\delta V_{\psi} = \psi_a\, \Bigl [ \eta\, DX^a - \g_b\eta\, \XX ab 
- \eta R^{ab}\, k_b - 3 \g_b\eta\, R^{abc}\, k_c\, \Bigr ]
\label{here}
\ee
where we made use of the Fierz identity
\be
\psi_a\th\, \th\g^{ab}\eta \, k_b =
-\psi_a\eta\, R^{ab}\, k_b + 3 \psi_a\g_b\eta\, R^{abc}\, k_c
\ee
ignoring longitudinal polarizations. From the longitudinal 
supergravity variations \eqn{dhab} and \eqn{dCabc} and the
$V_{h_+}$ and $V_{C_+}$ vertices of \eqn{Vh+} and \eqn{VC+}
we see that \eqn{here} reads $\delta V_{\psi} = -(V_{\delta h_+}
+V_{\delta C_+})/\sqrt{2}$ as expected.

For the more involved $\tilde\psi_a$ vertex let us analyze the resulting 
terms order by order in $\theta$ 
to keep the resulting expressions in a manageable size.
At zeroth order in $\theta$ one finds 
\bea
\delta V_{\tilde\psi} \Bigr |_{\theta^0} &=&
\tilde\psi_{(a}\g_{b)}\eta\, \Bigl[ DX^a\, DX^b -\XX a c\, \XX b c \Bigr ] 
+\ft 3 2 \, \tilde\psi_{[a}\g_{bc]}\eta\, DX^a\, \XX b c 
\nn\\&&
+\tilde\psi_a\eta\, DX^b\,
\XX b a
\eea
In the first line we can already recognize the $\theta$ independent 
terms of the transverse graviton and three-form vertex.

For the terms quadratic in $\theta$ let us first look at the term which
survives the particle reduction already discussed in \cite{GKG}
\bea
\delta \, \Bigl [ -2 \tilde\psi_a \,\g\cdot DX\, \th\, R^{ab}\, k_b\Bigr ]
&=&-2 \tilde\psi_{(a}\g_{b)}\eta\, DX^a\, R^{bc}\, k_c
-6 k_{[a}\tilde\psi_b\g_{cd]}\eta\, DX^a\, R^{bcd}\nn\\
&& -\tilde\psi_a\g_b\eta\, R^{ab}\, k\cdot DX - 3\tilde\psi_a\eta\, 
R^{abc}\, DX^b\, k_c
\label{one}
\eea
while for the remaining genuine membrane-like terms involving the
APD bracket one finds
\bea
\lefteqn{\delta \,  \Bigl [ 
8  \tilde\psi_a\gamma_b\theta\, \XX b c \, R^{cad}\, k_d 
+ \ft 5 3 \tilde\psi_a\gamma_{bc}\theta\, \XX b c R^{ad}\,k_d  
+ \ft 4 3 \tilde\psi_a\gamma_{bc}\theta\, \Bigl ( \XX a b\, R^{cd}}
\nn\\
\lefteqn{\quad
 + \XX c d R^{ab}
\Bigr ) k_d
+\ft 2 3 i \tilde\psi_a\,\Bigl ( \gamma_b\theta\, \{X^a,\theta\}\gamma^b\theta
-\theta\, \{X^a,\theta\}\theta \,\Bigr ) \Bigr ] }
\nn\\&=&
-6 \tilde\psi_{(a}\g_{b)}\eta\, \XX a c\, R^{bcd}\, k_d
+3 k_{[a}\, \tilde\psi_b \g_{cd]}\eta\, \XX a b R^{cd}
\nn\\&&
- \ft 12 k_{a}\, \tilde\psi_b \g_{cde}\eta\, \Bigl ( 
\XX a b R^{cde}-3 \XX bc R^{dea}-3 \XX e a R^{bcd}-3 \XX d e R^{abc}
\Bigr )
\nn\\&&
+3 \tilde\psi_a\g_b\eta\, \{k\cdot X, X^c\}\, R^{abc}
+i\tilde\psi_a\g_b\eta\, \th\g^b\{\th, X^a\}
+2 k_{[a}\tilde\psi_{b]}\eta\, \XX a c\, R^{bc} 
-i \tilde\psi_a\eta\, \{\th,X^a\} \nn\\
\label{two}
\eea
where we made use of several Fierz rearrangements,
in which one also invokes the physical state constraints 
\eqn{sugraconst} of the gravitino.
Now the first line of the variations in \eqn{one} and \eqn{two}
respectively together produce two of the three $\theta^3$ terms 
in the transverse graviton \eqn{Vh} and 3-from \eqn{VC} vertex.
Moreover the missing $\theta^3$-term of the three-from vertex
is actually given by the second line of the right hand side of
\eqn{two} as
\bea
\lefteqn{
\delta F_{abcd}\Bigl (- \ft 1{96} \XX e f \, \th\g^{abcdef}\th\, \Bigr )}\nn\\
&=&
-\ft 1{16} k_a\, \tilde\psi_b\g_{cd}\eta\, \XX e f \, \th\g^{abcdef}\th 
= -\ft 1 {48} k_a\, \tilde\psi_b\g_{abcdefg}\eta \, \XX cd \, \th
\g^{efg}\th \nn\\
&=& 
-\ft 1 2 k_{a}\, \tilde\psi_b \g_{cde}\eta\, \Bigl ( 
\XX a b R^{cde}-3 \XX bc R^{dea}-3 \XX e a R^{bcd}-3 \XX d e R^{abc}
\Bigr ) \, ,\nn\\
\eea
where we first dualized the gamma matrices and thereafter reduced
$k_a\, \tilde\psi_b\g_{abcdefg}\eta$ to expressions with three index
gamma matrices via the physical state constraints of the gravitino 
\eqn{sugraconst}. The missing term $-i\th\g^{(a}\{X^{b)},\th\}$
of the graviton vertex is found from \eqn{one} and \eqn{two}
by first partially integrating the first term of the last line of
\eqn{two}
\be
3 \tilde\psi_a\g_b\eta\, \{k\cdot X, X^c\}\, R^{abc}\, e^{-ik\cdot X} 
= \Bigl[ -\ft i 2 \tilde\psi_a\g_b\eta\, \th\g^{ab}\g^c\,\{\th,X^c\}
+i \tilde\psi_{[a}\g_{b]}\eta\, \th\g^a\{\th,X^b\}
\Bigr ] \, e^{-ik\cdot X} 
\label{three}
\ee
where we have also made use of the identity $\g^{abc}=\g^{ab}\g^c
- 2\g^{[a}\, \delta^{b]c}$. Now adding the second term of the last line of
\eqn{two} to \eqn{three} yields the desired symmetrized expression
$i \tilde\psi_{(a}\g_{b)}\eta \, \th\g^a\, \{\th,X^b\}$. What remains
to be shown, however, is that the first term in \eqn{three} cancels.
This is achieved by again partially integrating the first term of 
the second line of \eqn{one}
\be
 -\tilde\psi_a\g_b\eta\, R^{ab}\, k\cdot DX\, e^{-ik\cdot X} =
\ft i 2 \tilde\psi_a\g_b\eta\, \th\g^{ab}D\th\, e^{-ik\cdot X}
-k^-\, \tilde\psi_a\g_b\eta\, R^{ab}\, e^{-ik\cdot X}
\ee
which thus cancels the first term in \eqn{three} upon using the equation
of motions for $\th$ of \eqn{eomth}.

Putting it all together we arrive at the final result
\bea
\delta V_\Psi &=& -V_{\delta h} -V_{\delta C} 
-\ft{1}{\sqrt 2}\, V_{\delta h_+}
- \ft 1 {\sqrt{2}} V_{\delta C_+}
\nn\\
&& + \tilde\psi_a\eta\, DX^b\XX b a - 2k_{[a}\, \tilde\psi_{b]}\eta
\, R^{ac}\, \XX b c - i \tilde \psi_a\eta\, \th\{\th,X^a\}\nn\\&&
-3 \tilde\psi_a\eta\, R^{abc}\, DX^b\, k_c
+\ft 4 3 \tilde\psi_a\eta\, R^{abc}\, R^{bd}\, k_c\, k_d
-k^-\, \tilde\psi_a\g_b\eta\, R^{ab} \, .
\eea
The remaining terms in the second and third line are associated 
with the longitudinal parts of the vertex operators whose 
polarization components vanish by our gauge choices: for instance, 
it is easy to see that the terms multiplying $\tilde\psi_a \eta$ 
arise in the $d=11$ supersymmetry variation of $C_{a+-}$ and 
therefore belong to the vertex operator for $C_{a+-}$ which may now be
read off as
\bea
V_{C_{+-}}&=& C_{a+-}\, \Bigl[ DX^b\, \XX{b}{a}- i\theta\, \{\theta, X^a\}
-3 R^{abc}\, DX^b\, k_c \nn\\&&
+\XX{a}{c}\, R^{cd}\, k_d -\ft i 2 \, \theta
\gamma^{ac}\, \{\theta, X^c\}
+\ft 4 3 R^{abc}\, R^{bd}\, k_c\, k_d
\, \Bigr ] \, e^{-i\, k\cdot X}\, . 
\label{VC+-}
\eea
Also, the gauge invariance of $V_{C_{+-}}$ may be checked easily.
It is important to realize that these longitudinal operators do 
appear in the variations even if their polarizations have been 
set to zero. Compensating gauge transformations are not relevant 
here, as the vertex operators are inert under these transformations.

\subsection{Nonlinear Supersymmetry}

The non-linear supersymmetry transformations on the vertex operators give  
further consistency checks. They constitute the $\epsilon$ 
dependent transformations as
given in (\ref{susytrafos}). We restate them here, and denote 
the transformations
as $\tilde{\d}$. The APD brackets play a major role in the 
non-linear supersymmetry 
of the supermembrane coordinates and mark a difference from the 
superparticle. This
also makes these transformations non-trivial. 
\bea
\tilde{\d} X^a &=& -2 \epsilon\gamma^a\theta \nn\\
\tilde{\d} \theta &=& i\, DX\cdot \gamma\, \epsilon -\ft i 2 \{X^a,X^b\}\, 
\gamma_{ab}\epsilon \nn\\
\tilde{\d} DX^a&=& -2 \epsilon \gamma^a D\theta + 2\{\epsilon\theta, X^a\}
\label{inlsusytrafos}
\eea
The corresponding transformations for supergravity wave functions are 
\cite{GKG}\footnote{We 
stress once more that the compensating gauge transformations
considered in \cite{GKG}, which are singular in $k^+$, are redundant
as they vanish when contracted 
into the corresponding vertex operators by the gauge invariance of 
the latter.}
\bea
\tilde {\d}h_{ab}&= &\e \g_{\left(a\right.}\psi_{\left.b\right)} \\
\tilde{\d}\psi_a &=& -k_{\left[+\right.}h_{\left. b\right] a}\g^b\e 
\label{nlnpsi}\\
\tilde{\d}\tilde{\psi_a}&=& k_{\left[c\right.}h_{\left.b\right] a}
\g^{cb}\e +\frac{1}{72}\left(
\g_a^{bcde}F_{bcde} + 24 \g^{bcd}k_bC_{acd} + 4\g^{bcd}k_{a}C_{bcd}
\right)\e\label{nlnpsitilde}\\
\tilde{\d}C_{abc}&=& \frac{3}{2}\e \g_{\left[ab\right.}\psi_{\left. c\right]} 
\, .
\eea
We have quoted the transformations of the transverse components only, 
as we shall only need those in the following discussion. 
In fact, here we present only the transformation of the 
transverse graviton vertex, and show that  
 the terms remarkably combine to give the expected gravitino
vertices and total derivative terms. The graviton vertex (\ref{Vh}) 
under non-linear supersymmetry gives:
\bea
\tilde{\d} V_h &= & h_{a b}\left[ 4(-\e\g^aD\th + \{\e\th,X^a\})DX^b 
+ 4\left(\{\e\g^a\th , X^c\} 
+ \{X^a,\e\g^c\th\}\right)~\{X^b,X^c\}\right. \nn \\&& 
-2\{X^b,DX^c\}\e\g^c\g^a\th
-\{X^b,\{X^e,X^f\}\}\e\g^{ef}\g^a\th  \nn \\
&&-i DX^a\left( DX^e\e \g^e + \frac12 \{X^e,X^f\}
\e\g^{ef}\right)\g^{bc}\th k_c \nn \\
&&-\left. i\{X^a,X^c\}\left(DX^e\e \g^e + 
\frac12\{X^e,X^f\}\e\g^{ef}\right)\g^{bcd}k_d\th\right]e^{-ik\cdot X}\nn \\
&&+i\tilde{\d}\theta\g^a\{X^b, e^{-ik \cdot X}\}\theta + 2i \e k\cdot\g 
\th~ V_h
\label{var} 
\eea
We ignore terms of order $\th^3$ for simplicity at present.
The terms in the third 
and fourth line of the above equation yield most of the relevant terms.
They can be combined as: 
\be 
-i \left(DX^e\e\g^e + \frac12 \XX ef \e \g^{ef}\right)\g^{bd}
k_d\left(DX^a + \g^c\XX ca\right)\th
-i h_{ab}\{X^b,e^{-i k\cdot X}\}\tilde{\d}\th\g^a\th
\label{com}
\ee
After a few manipulations in which we commute the $\g^{bd}$ to the 
left to contract it with
$\e h_{ab}$, we get the following from (\ref{com}) :
\bea 
&& i h_{a\left[b\right.}k_{d\left. \right]}\epsilon\g^{db}\left(DX 
\cdot \g + \frac12 \XX ef 
\g^{ef}\right)\left(DX^a + \g^c\XX ca\right)\th\nn\\
&&+2i h_{ab}k^{-}\epsilon\g^b \left( DX^a + \g^c\XX ca\right)\th 
-2h_{ab}\e \g^aD\left[\left(DX^b + \g^c\XX ca\right)\th 
e^{-ik\cdot X}\right]\nn\\
&&+h_{ab}\left[ 4\e \g^a D\th DX^b - 2\e \g^a\g^c\{X^b,DX^c\} 
-4\{\e\g^a\th,X^c\}\XX bc \right.\nn\\
&&-4\{X^a,\e\g^c\th\}\XX bc -2i \e k\cdot\g\left(DX^a DX^b 
-\XX ac \XX bc\right)\th \nn \\
&&\left. + 2\{X^a,\XX cb\}\e\g^c\th 
  - 2\{X^e,\XX ac\} \e \g^{ebc}\th\right]e^{-i k\cdot X} \nn \\
&&+4 h_{ab}\e\th DX^a\{e^{-ik\cdot X},X^b\}
-i h_{ab}\{X^b,e^{-i k\cdot X}\}\tilde{\d}{\th}\g^a\th - \partial_r W'^r
\label{exp}
\eea
Here $\partial_r W'^r$ comes from the partial integration of terms  
proportional to $\{X^a,e^{-ik\cdot X}\}$, where $W'^r= 2\e^{rs}h_{ab}
\left( \e \g^b\g^e\th \partial_s X^e DX^a +
\e\g^e\g^b\g^c\th \partial_s X^e\XX ca \right)e^{-i k \cdot X}$. 
Unlike the superparticle case considered in \cite{GKG}, 
where $\dot{\th}=\ddot X=0$
the total derivative term in the second line of (\ref{exp}) involves 
quite a few terms proportional to $D\th, D^2X$
and $\{DX,X\}$ not present for the superparticle, and it is remarkable 
that the non-linear supersymmetry variation of the supermembrane
yields all the required derivatives. We use the equations of motion given in
(\ref{eomomega}) extensively in the above and in particular, we take $D^2X^a 
+ \{X^c,\XX ac\}=0$ at
this order in $\theta$. Substituting (\ref{exp}), in (\ref{var}), we get
\bea
\tilde{\d}V_{h}&= &\tilde{\d}\psi_a\left(DX^a + \g^c\XX ca\right)\th\nn\\ 
&&+\tilde{\d}\tilde{\psi}_a\left(DX \cdot \g + \frac12 \XX bc \g^{bc}\right)
\left(DX^a + \g^d\XX da\right)\th
\nn \\
&&+ h_{ab}\e\left[ -\{X^a,\XX ef\}\g^{efb} - 2\{X^e,\XX af\}\g^{ebf}\right]\th\nn \\
&&-\e DW^0 - \e \partial_rW^r
\label{fin}
\eea
The term in the
third line of (\ref{fin}) is easily seen to vanish by Jacobi identity. 
The gravitino vertex given in (\ref{VPsi}) is
also clearly recovered to this order in $\theta$ (from (\ref{nlnpsi}) 
and (\ref{nlnpsitilde}), $\tilde{\d}\psi~\propto ~k^-h_{ab}\g^b\e,
~\tilde{\d}\tilde{\psi}~\propto~ k_{\left[d\right.}
h_{\left.b\right]a}\g^{bd}\e$). 
 Also,  $W^r= W'^r -4 h_{ab}\e^{rs}\partial_s X^aDX^b\e\th$. The functions 
\bea
W^0 &= &2h_{ab}\g^a(DX^b + \g^c\XX cb)\th e^{-i k\cdot X}\; \mbox{and} \nn \\
 W^r &= &2\e^{rs}h_{ab}\e\g^e\g^a\partial_s X^e \left(DX^b + \g^c\XX cb\right)
\th e^{-i k\cdot X}
\eea
are also expected to obey certain transformation properties under 
supersymmetry as given for the
superparticle in {\cite{GKG}}. However, we have
not checked for them, and it shall be interesting to investigate them in 
the future.

The variation of the graviton vertex into the gravitino vertex to 
order $\theta^3$
involves more tedious computations, and we refrain from checking for all 
the terms.
However, it is easy to see that the following terms arises in the variation
\bea
h_{a b}\left(4\e \g^a D\th R^{bc}k_c + 2\e \g^a\th \th\g^{bc}D\th 
k_c - 4 i \e \g^a \th DX\cdot k R^{bc}k_c\right)e^{-ik\cdot X}  
\nn \eea
This term can be combined into a total derivative and the 
variation of $\psi$, as
\bea
\tilde{\delta}\psi_a (-2\th R^{ac}\, k_c)e^{-ik\cdot X} + 
D\left(4h_{ab} \e \g^a\th R^{bc} \, k_c \, e^{-ik\cdot X}\right) \nn
\eea
Thus the vertex for $\psi_a$ is recovered to all orders in $\theta$ here. 

\sect{Applications to M(atrix) Theory}

Our results immediately imply two important applications to matrix 
theory. Firstly, we now have the lagrangian for the light cone 
supermembrane in a weak background, as the vertex operators represent 
nothing but the linear coupling of the background fields to the 
supermembrane coordinates $X^a$ and $\th$. Hence
\be
{\cal L}_{\mbox{\footnotesize weak}}= {\cal L} + V_{h(X)} + V_{h_+(X)} 
+ V_{h_{++}(X)} + V_{C(X)} + V_{C_+(X)} + V_{\Psi(X)} + V_{\Psi_+(X)}
\label{m2weak}
\ee
where ${\cal L}$ denotes the supermembrane lagrangian in 
flat space \eqn{m2lagrangian} and where one writes the vertex operators
of \eqn{Vh}-\eqn{VPsi-} in configuration space, e.g.
\be
V_{h_+(X)}=  -2 \, \left(DX^a - R^{ab} \frac{\partial}{\partial X^b}
     \right)\, h_{a+}(X)
\ee
for the linear coupling to the background field $h_{+a}(X)$. 
We stress that we now know this action
to {\it all} orders in $\theta$, which is to be contrasted with
the results on the covariant supermembrane in general
background fields \cite{dWPP} where the action was derived to all
orders in the background fields, but only up to order $\theta^2$ in
the membrane fermions\footnote{In \cite{dWPPS} the covariant membrane 
action for the $AdS_4\times S_7$ and $AdS_7\times S_4$ backgrounds
was obtained to all orders in $\th$. It would be interesting 
to compare these results to ours by taking the action of 
\cite{dWPPS} to the light-cone.}. Clearly our results
immediately carry over to matrix theory: one needs only repeat the usual 
matrix model regulation \cite{GH,dWHN} of the light-cone supermembrane
using the prescription of \eqn{mtrans}
\be
X^a(\tau, \s_1,\s_2) \rightarrow {\bf X}^a_{mn}(\tau)\qquad
\th^\a(\tau, \s_1,\s_2) \rightarrow {\theta}^\a_{mn}(\tau)
\ee
with $n,m=1,\ldots, N$ labeling the entries of the $N\times N$
hermitian matrices ${\bf X}^a$ and $\theta^\a$. Moreover the APD
Lie bracket gets replaced by a matrix commutator
$\{.\, , .\} \rightarrow i\, [.\, , .]$.
The only subtlety in replacing world-space integrals by traces
occurs in expressions of higher than second order in the matrices 
$D{\bf X}^a$, $\th^a$, $[{\bf X}^a, {\bf X}^b]$, 
$[\th,{\bf X}^b]$ and $\exp[-i\, k\cdot {\bf X}]$, where we must
deal with ordering ambiguities under the trace. However, in
order to maintain the defining transformation properties 
of the vertex operators under gauge symmetry and supersymmetry 
discussed in section 3 for the matrix theory regulation it is 
sufficient to replace the world-space integral 
by a {\it symmetrized trace}, i.e. 
\be
\ft 1{4\pi}\int d^2\s\, (\ldots )
\rightarrow \ft 1 N \, \mbox{STr}[\ldots ]
\ee
where it is understood that the symmetrization in the trace 
is to be performed over the set of matrices 
($D{\bf X}^a$, $\th^a$, $[{\bf X}^a,{\bf X}^b]$, $[\th,{\bf X}^b]$,
$\exp[-i\, k\cdot {\bf X}]$). The vertex operators obtained in
this way may be compared to the results of Taylor and Van Raamsdonk 
\cite{TVR}, who derived certain expressions for the energy-momentum 
tensor, the membrane current 
and supercurrent of matrix theory up to 
quadratic order in $\theta$ and (partially) up to linear order 
in transverse space derivatives $\partial_a$ (related to the $k_a$ in 
the momentum picture). Their results are based on a one-loop 
matrix theory computation for general block diagonal matrix 
backgrounds. Happily, we find agreement with their results to the 
order that they have computed\footnote{But there seems to be a mismatch in 
one order $\theta^2$ term in the three-form vertex (membrane current).}. 
However, there are additional operators in the matrix theory 
picture of \cite{TVR} coupling to the background fields 
$h_{a-}$, $h_{--}$, $C_{ab-}$, $C_{a+-}$, $\Psi_-$, 
which we have gauged to zero\footnote{Our result for $V_{C_{+-}}$ of
\eqn{VC+-} also agrees with \cite{TVR} to the order that they have
computed.}. 

Besides the background field matrix theory action obtained above,
we believe that another interesting application of the supermembrane
vertex operators lies in a new definition of scattering amplitudes
in matrix theory. Conventionally these are evaluated by computing
an effective background field action through
a fluctuation expansion around diagonal matrix backgrounds obeying
the classical equations of motion. The obtained effective action is then
Fourier transformed and sandwiched between polarization states in order
to obtain genuine S-matrix element in
momentum space \cite{SPW}. This approach only allows for the
computation of amplitudes in the eikonal (zero momentum transfer) limit.
Moreover it completely neglects bound state effects as it models the
complicated matrix theory ground state by semiclassical diagonal
matrix configurations. 

With the matrix theory vertex operators at hand a much more natural
definition of $n$-particle scattering amplitudes is given by the path integral
\bea
{\cal A}_{{\cal H}_1\ldots {\cal H}_n} &=& 
\Bigl\langle \prod_{j=1}^n \int d \tau_j\, \mbox{STr} \,\Bigl (
 V_{{\cal H}_j}[{\bf X}^a(\tau_j),\th(\tau_j)]\, \Bigr ) 
\Bigr\rangle \nn\\&=&
\int {\cal D} [{\bf X}^a,\th] \,
\prod_{j=1}^n \int d \tau_j\, \mbox{STr}\, \Bigl (
V_{{\cal H}_j}[{\bf X}^a(\tau_j),\th(\tau_j)] \,\Bigr )\,
e^{i\, S_{MT}[{\bf X},\th]}
\label{AH}
\eea
where ${\cal H}_j$ denotes the polarization and momentum of the $j$'th
particle. It remains to be seen whether a (perturbative) evaluation of
\eqn{AH} makes sense, because in contrast to the superstring or the
superparticle we are now dealing with the computation of expectation
values of composite operators in an {\it interacting} theory. However,
the definition \eqn{AH} overcomes the restriction to the eikonal sector
of the conventional approach, it should include large $N$ and bound 
state effects and manifestly obeys supersymmetric Ward identities, which
is far from obvious in the conventional approach. Also at least
for the further reduction to the zero dimensional IKKT matrix model
of IIB theory \cite{IKKT} a numerical evaluation of scattering 
amplitudes along the lines of \cite{NKS} may now become feasible.

\sect{Outlook}
In this paper we have demonstrated that the supermembrane and the 
associated supersymmetric APD gauge theory contain the type II
superstrings and the matrix model not only at the level of
the action, but also at the level of the vertex operators expected
to describe various physical processes. Although a full quantum
treatment of the supermembrane or the equivalent supersymmetric
APD gauge theory still seems difficult, we can now explore the 
theory much further {\it at the dynamical level} by matching it
in the appropriate domains with the simpler subtheories that must 
be consistently contained in it. In particular, we have in mind the 
following comparisons:

\begin{itemize}
\item The $d=11$ superparticle reduction has been used in \cite{GGV}
   to determine the non-perturbative contributions to the $R^4$ corrections 
   to the effective string action in terms of non-holomorphic 
   Eisenstein series (also computed in \cite{RT}). 
   Remarkably, this calculation makes use of only a single term 
   in the graviton vertex (\ref{Vh}), namely the zero mode of 
   $h_{ab}R^{ac}R^{bd}k_c k_d$ (the coefficient of the fermionic 
   quadrilinear is easily seen to coincide with the linearized 
   Riemann tensor). The resulting infinite sum over D instanton
   contributions can be alternatively viewed as a sum over BPS 
   multiplets \cite{dWL}. However, in order to arrive at a finite result
   a divergent term must be discarded ``by hand'' \cite{GKG,dWL}. 
   This infinity should disappear when the M theory 
   degrees of freedom are properly taken into account.
\item As already pointed out in the foregoing section, the matrix
   theory vertex operators afford an entirely novel approach
   to the computation of scattering amplitudes. In particular,
   it should now be possible to determine these beyond the eikonal
   regime. The computation of $R^4$ corrections within the framework
   of matrix theory will have to be re-examined.
\item Superstring amplitudes should emerge in the superstring limit. 
   While the matrix theory scattering amplitude (\ref{AH}) is one
   way to approximate the APD gauge theory path integral
   \be
   \int {\cal D} [X^a,\th] \prod_j \int d\tau_j d^2\s_j 
   V_{{\cal H}_j}[X^a(\tau_j,\s_j),\th(\tau_j, \s_j)] e^{i\, S_{APD}}
   \ee
   the superstring amplitudes are obtained in a very different limit 
   of the same expression. In that approximation one looks at the 
   regions where the membrane degenerates into a multi-string
   configuration, and the vertex operator insertions reduce to
   superstring vertex operators, as we have shown. In this way, 
   one should also be able to recover {\it multi-string vertex
   operators} (see e.g. \cite{NW} and references therein) from
   the quantum supermembrane.

\end{itemize}

Finally, we would like to emphasize once more the intrinsic multi-particle
nature of the theory, which is the main conceptual difference between 
supermembrane and superstring theory: it appears to be impossible to
tackle supermembrane theory by first defining one-particle excitations,
and subsequently second-quantizing it so as to obtain its multi-particle 
states. Therefore, unlike for superstring theory, the conventional
Fock space quantization breaks down. An interesting consequence of
this conclusion is that there should not exist any analog of the vertex 
operators corresponding to excited (massive) string states.

\vspace{6mm}
\noindent
{\bf Acknowledgement}

We are grateful to M.B. Green for correspondence. H.N. would like to 
thank the CERN Theory Division, where this work was first presented, 
for hospitality.

\end{document}